\documentclass[journal]{IEEEtran}
 
\usepackage{cite}
\usepackage{amsmath,amssymb,amsfonts}
\usepackage{algorithmic}
\usepackage{algorithm}
\usepackage{graphicx}
\usepackage{textcomp}
\usepackage{xcolor}
\usepackage{booktabs}
\usepackage{multirow}
\usepackage{array}
\usepackage{colortbl}
\usepackage[hidelinks]{hyperref}
\usepackage{url}
\usepackage{tabularx}
\definecolor{headerblue}{RGB}{0, 112, 192}
\definecolor{orange80}{RGB}{255,165,0}
\definecolor{lightblue}{rgb}{0.85, 0.92, 1}
\ifCLASSINFOpdf
\else
\fi
\begin{document}
%
% paper title
% Titles are generally capitalized except for words such as a, an, and, as,
% at, but, by, for, in, nor, of, on, or, the, to and up, which are usually
% not capitalized unless they are the first or last word of the title.
% Linebreaks \\ can be used within to get better formatting as desired.
% Do not put math or special symbols in the title.
\title{PPO guided Agentic Pipeline for Adaptive Prompt Selection and Test Case Generation}
%
%
% author names and IEEE memberships
% note positions of commas and nonbreaking spaces ( ~ ) LaTeX will not break
% a structure at a ~ so this keeps an author's name from being broken across
% two lines.
% use \thanks{} to gain access to the first footnote area
% a separate \thanks must be used for each paragraph as LaTeX2e's \thanks
% was not built to handle multiple paragraphs
%

\author{Gourisetty Venkata Sai Koushik,~\IEEEmembership{Department of Computer Science and Engineering, SRM University AP}\\
        Dama Aditya,~\IEEEmembership{Department of Computer Science and Engineering, SRM University AP}\\
         Mahankali Harish Sai,~\IEEEmembership{Department of Computer Science and Engineering, SRM University AP}
        Peddi Siddarhta,~\IEEEmembership{Department of Computer Science and Engineering, SRM University AP}\\
        Shadab Ahmad,~\IEEEmembership{Department of Computer Science and Engineering, SRM University AP}\\
        and~Vivek~Yelleti,~\IEEEmembership{Department of Computer Science and Engineering, SRM University AP}% <-this % stops a space
\thanks{All authors are with the Department
of Computer Science and Engineering, SRM University AP, India.}
\thanks{Corresponding Author: Vivek Yelleti, was with the Department
of Computer Science and Engineering, SRM University AP,
India, e-mail: vivek.yelleti@gmail.com.}% <-this % stops a space
}

% note the % following the last \IEEEmembership and also \thanks - 
% these prevent an unwanted space from occurring between the last author name
% and the end of the author line. i.e., if you had this:
% 
% \author{....lastname \thanks{...} \thanks{...} }
%                     ^------------^------------^----Do not want these spaces!
%
% a space would be appended to the last name and could cause every name on that
% line to be shifted left slightly. This is one of those "LaTeX things". For
% instance, "\textbf{A} \textbf{B}" will typeset as "A B" not "AB". To get
% "AB" then you have to do: "\textbf{A}\textbf{B}"
% \thanks is no different in this regard, so shield the last } of each \thanks
% that ends a line with a % and do not let a space in before the next \thanks.
% Spaces after \IEEEmembership other than the last one are OK (and needed) as
% you are supposed to have spaces between the names. For what it is worth,
% this is a minor point as most people would not even notice if the said evil
% space somehow managed to creep in.

% The paper headers
\markboth{Journal of \LaTeX\ Class Files,~Vol.~14, No.~8, August~2015}%
{Shell \MakeLowercase{\textit{et al.}}: Bare Demo of IEEEtran.cls for IEEE Journals}
% The only time the second header will appear is for the odd numbered pages
% after the title page when using the twoside option.
% 
% *** Note that you probably will NOT want to include the author's ***
% *** name in the headers of peer review papers.                   ***
% You can use \ifCLASSOPTIONpeerreview for conditional compilation here if
% you desire.

% If you want to put a publisher's ID mark on the page you can do it like
% this:
%\IEEEpubid{0000--0000/00\$00.00~\copyright~2015 IEEE}
% Remember, if you use this you must call \IEEEpubidadjcol in the second
% column for its text to clear the IEEEpubid mark.

% use for special paper notices
%\IEEEspecialpapernotice{(Invited Paper)}

% make the title area
\maketitle

% As a general rule, do not put math, special symbols or citations
% in the abstract or keywords.
\begin{abstract}
Developing effective test cases capable of thoroughly exercising large-scale software systems is inherently difficult, especially if such systems have voluminous, complex, and deeply nested source codes. In this work, we present a novel approach for generating test cases using a reinforcement learning-driven agentic framework where Proximal Policy Optimization (PPO) is coupled with an LLM engine to guide prompt selection during test generation. Our approach consists of two phases. In Phase I, the ToT-guided optimization agent partitions and minimizes the source code by removing redundancies without changing the functional behavior of the source code. In Phase II, a PPO-based policy network is trained to solve the problem of selecting prompts among eight different prompting techniques, such as Boundary Value Analysis, Random Fuzzing, etc., based on the inputted 11-dimensional state vector representing the source code complexity metrics and live coverage metrics to direct the LLM engine towards exploring unvisited paths in the program. The PPO agent receives rewards based on a combination of increases in line and branch coverages, penalties for unexplored branches, and rewards for reducing source code length. From experiments conducted on twenty benchmark programs, it is evident that the proposed approach, PPO-LLM, outperforms CBMC, kS-LLM, and kS-LLM++ in terms of branch and line coverage in almost all cases, for various loop bound values ranging from BOUND~1 to BOUND~2000. While at BOUND~1, the coverage of branches is 100\% using PPO-LLM on the PALS suite, in comparison, it is around 86.8\% using kS-LLM++. This confirms that adaptive prompt selection driven by PPO substantially outperforms static prompting strategies on PALS type programs.
\end{abstract}

% Note that keywords are not normally used for peerreview papers.
\begin{IEEEkeywords}
Reinforcement Learning, Proximal Policy Optimization, Large Language Models, Automated Test Case Generation, Branch Coverage, Line Coverage, Code Optimization, Tree-of-Thought, Markov Decision Process
\end{IEEEkeywords}

% For peer review papers, you can put extra information on the cover
% page as needed:
% \ifCLASSOPTIONpeerreview
% \begin{center} \bfseries EDICS Category: 3-BBND \end{center}
% \fi
%
% For peerreview papers, this IEEEtran command inserts a page break and
% creates the second title. It will be ignored for other modes.
\IEEEpeerreviewmaketitle

\section{Introduction}

Reliability of software is no longer a mere theoretical objective; instead, it is a real engineering necessity that gets harder and harder to satisfy as today's complex systems continue to grow larger and more tightly coupled. The mismatch between the programmer's expectations regarding the behavior of their code and what the code actually behaves like in response to malicious or unexpected inputs has led to anything from simple disruptions in services to major accidents in crucial systems. ATCG remains a critical part of solving this problem, because well-crafted test cases can help explore the execution paths that are too hard for humans to discover.

Traditional ATCG algorithms have had years to develop and improve upon themselves. Symbolic execution tries to account for every possible input by performing symbolic computations rather than concrete ones across the control flow graph. CBMC, on the other hand, reduces the problem of reachability to SAT and then leverages an existing SAT solver. Both methods are theoretically valid approaches, but they suffer from the same basic issue: state explosion. The more the size of the program increases, either in lines of code, branches, or loops, or due to complex data structures, the amount of resources necessary to analyze every single execution path increases exponentially until it becomes impractical due to hardware limitations. This applies in particular to CBMC on real-life files with hundreds to thousands of lines.

Recently, large language models have been found to be an attractive choice of mechanism to generate tests. However, unlike the symbolic approaches, LLMs are not based on enumerating program states, but leverage their understanding of what inputs are good for finding bugs and exploring uncharted territory of the program obtained by training on a huge amount of code and natural language data. There are several papers demonstrating how a well-designed prompting can yield high-quality tests in case of a reasonably sized program. However, one word in the above sentence needs a special mention: well-designed. The efficiency of LLM-generated tests strongly depends on the design of the prompt, and, for example, if a prompt focuses on boundary conditions, it is expected to yield more relevant tests. At the same time, using only one prompting pattern will probably be inefficient in case of multiple stages of increasing coverage, since the areas uncovered by the testing process will change with each step.

This is exactly where prompt selection comes into play as a critical design choice. Here is what actually goes on: the prompt designed for handling boundary values does well during the initial stages of the generation process, where simple branches have yet to be covered. But as simple branches become fully covered, uncovered gaps tend to become deeper and more unusual maybe even conditional upon integer overflow, environmental state changes or some rare error conditions. A prompt designed for boundary values will have very little insight into such paths. Running the same prompt against the model time after time only drains the generation budget without advancing coverage further. What needs to happen next is switching prompts to those focusing on error paths, loop boundaries or similar. But since prompts are hardcoded, there is no way for the system to detect this plateau and do anything about it. Prompt selection should therefore not be viewed as a formality, it is the key which determines how the model's attention will be focused and it is as important as any capability of the underlying model.

kS-LLM, the earlier framework, made a valuable contribution by integrating a k-step cache with LLM-generated responses, while its later version, kS-LLM++, further enhanced the approach with Tree-of-Thought code optimization in a two-stage pipeline. Even with these advancements, both models follow the same static prompting paradigm that fails to adjust itself according to coverage. In a scenario where coverage gets stuck at a relatively low point, there is no way for the system to detect this phenomenon and make appropriate changes to its strategy.

This is where the proposed approach comes into play, which considers the task of selecting the template as a Markov Decision Process and uses Proximal Policy Optimization to solve it by selecting the best template based on the observed code complexity properties and coverage statistics. This approach enables a smooth transition from exploration mode, when the system generates various inputs belonging to different categories, to exploitation mode, when the system focuses on hammering the uncovered branches using tailored strategies.

The central contributions of this work are:
\begin{itemize}
    \item \textbf{MDP formulation for prompt selection:} The selection of prompt templates is framed as a sequential decision-making process, allowing us to optimize the selection strategy using reinforcement learning techniques, rather than relying on manual rules.

    \item \textbf{LLM prompt policy network using PPO framework:} Our policy network architecture involves the use of a two-layer multilayer perceptron (11$\rightarrow$32$\rightarrow$8). The neural net is trained with the objective function of PPO with the surrogate loss clipped, GAE, and entropy regularization.

    \item \textbf{Eight Prompt Template Library for Specific Prompt Strategies:} We design and implement a library of eight specialized prompting strategies: Boundary Value Analysis, Branch Coverage Explorer, Edge Case Hunter, Error Path Exerciser, Loop Boundary Tester, Data Type Stress, Path Coverage Maximizer and Random Fuzzing.

    \item \textbf{Rewards shaping that integrates coverage and code quality:} The definition of a composite reward, where we factor in gain in line and branch coverage, penalize unvisited branches based on the proportion visited, and provide a bonus for working with heavily compressed code.

    \item \textbf{Empirical Evaluation on 160 Benchmarks:} The approach is evaluated using 20 benchmarks selected from different benchmark suites that cover eight bound levels, showing consistent empirical improvement over CBMC, kS-LLM, and kS-LLM++ in many cases.
\end{itemize}

The rest of the paper is organized as follows: In Section II, we cover the related works. The proposed methodology is presented in Section III, and the dataset and metrics are described in Section IV. Results are presented and discussed in Section V. The conclusion is presented in Section VI.

% ============================================================
\section{Related Work}
Algarsamy et al. \cite{ALAGARSAMY2024107565} utilized the assertion method to generate the unit test cases and named it as A3Test. Further, they also included a method to consider the domain knowledge while generating the test cases. Ni et al. \cite{11229491} proposed CasModaTest that works in the staged approach while generating the test cases as follows: (i) initially, the provided code is divided into multiple cascaded ones; (ii) later, test prefix generation is invoked, followed by (iii) invoking the test oracle generation. Arya et al. \cite{11264985} proposed a method where they utilized NLP techniques and reinforcement learning methods to generate the unit test cases. Kikuma et al. \cite{10.1145/3368926.3369679} proposed a machine learning-based method to extract the homogeneous test cases which are completely independent of skills and mimic the style of an expert who is obliged to write the test cases. It analyzes the requirements specification document and then, using past development knowledge, its method generates homogeneous test cases. Tuncali et al. \cite{8500421} proposed a method to generate adversarial test cases to authenticate the efficacy of autonomous vehicles. In another work, Tuncali et al. \cite{8911483} proposed the utilization of a software requirements document to generate the test cases, which will ultimately serve as the adversarial test cases. Esnaashari et al. \cite{esnaashari2021automation} proposed a memetic algorithm, where the test cases are generated by the genetic algorithm. However, to improve the exploitation and to maintain diversity, they introduced reinforcement learning as the local search method. Lafi et al. \cite{9491761} employed machine learning algorithms to analyze the given software requirements document and then generate the test cases. Corradini et al. \cite{10.1145/3691620.3695511} proposed Deeprest, a framework where they employed a reinforcement learning algorithm to generate the test cases for Rest APIs. Koo et al. \cite{8730198} proposed PySE, which analyzes and codes and then derives the information from the branch execution. Later, this policy keeps getting updated as per the received in information from the underlying reinforcement learning agent. Tufano et al. \cite{tufano2020unit} proposed transformer-based unit test case generation. Additionally, they improved the performance by introducing the focal context in the learning method. With this method, transformer gave more importance to the relevant token thereby helped it to achieve more related unit test cases. Samuel et al. \cite{samuel2008automatic} proposed the method to generate the test cases by analyzing the UML diagrams. Liu et al. \cite{10.1145/3575693.3575707} proposed the NNSmith tool, which utilizes the fuzzing approach to analyze the deep learning compilers and then generate the test cases. Their novel fuzzing approach helped to generate the diverse and valid test cases. Yue et al. \cite{10.1145/2771783.2771799} proposed a Toucan4Test tool and considered the specification from the testing team, and analyzed these documents to generate the test cases. They validated their approach over the two industrial case studies and examined 30 automatically generated TCSs. Seqerloo et al. \cite{yazdani2019automatic} considered analyzing business models to generate test cases. They converted the business model into state graphs and then utilized them to generate the test cases, thereby increasing the diversity and relevance as per the business problem. Steenhoek et al. \cite{11026897} proposed method that works in a two-stage approach where initially, LLMs generate the test cases. Later, the reinforcement learning algorithm is invoked to reduce the test smells in the generated test cases by optimizing five coding quality controls and reduced the test smells. Sami et al. \cite{sami2024tool} proposed LLM driven based on prompt engineering for the test case generation. However, it analyzes the requirements document to accomplish this task. Mathur et al. \cite{10112971} employed T5 and GPT-3 to generate the test cases. These two methods embed the context-based meaning, which helps to identify the relevant keywords and generate test cases based on them. Zhang et al. \cite{zhang2024testbench} evaluated LLM models such as CodeLlama-13b, GPT-3.5, and GPT-4  for generating the class-level test cases. Their results demonstrated that the smaller models tend to be more sensitive to the noise and are contained within the full context. 
\vspace{1.5em}
A three-stage approach for creating test cases for dynamic programming languages such as Python was proposed by Yang et al. \cite{yang2025llm}. Their three-step procedure is as follows: (i) the parameter type is first examined; (ii) significant data is gathered during the mutations; and (iii) LLMs are used to fix the test cases that are produced. ChatUniTest, which uses a generate-validate-repair approach, was introduced by Chen et al. \cite{10.1145/3663529.3663801}. In this example, LLMs are used both while creating unit test cases and in these three distinct processes. A methodology to align test case generation with business needs was proposed by Hasan et al. \cite{11025799}. They assessed a number of models, and the findings showed that the Llama and Mistral models performed better than the others. Kurian \cite{10172742} developed an LLM-driven architecture for the Scade programming language, a contemporary programming language used for extremely important systems like as railroads. Lukasczyk and Fraser \cite{10.1145/3510454.3516829} suggested an effective system made up of independent agents. Xue et al. \cite{xue2024llm4fin} finetuned the pretrained model and developed a domain-dependent framework, and named it LLM4Fin. This model helped to generate the test cases, and then the performance of this model is validated through real-world stock trading. Their empirical analysis demonstrated that their LLM4Fin outperformed ChatGPT and other LLM models. Korraplu et al.\cite{korraprolu2025test} performed benchmark study over the popular six different LLMs. Their results indicated that the Gemini turned out to be the best in various metrics. Deng et al. \cite{deng2024llm} proposed method to construct the prompts by considering and analyzing the register transfer level behavior. The quality of the prompt dictates the performance of the LLM model. Gao et al. \cite{gao2025prompt} proposed an optimal way to design the user-defined prompts for automated test case generation. Schafer et al. \cite{10329992} proposed an approach wherein the user initially uploads the API of the project. Thereafter, the prompts are  designed automatically based on the signature and type of the functions as per the documentation provided by the developer. Koziolek et al. \cite{10711016} employed symbolic execution and search-based techniques to generate test cases. They validated the performance of their model over the PLC and DLC control logics.

\vspace{1.5cm}
% ============================================================

\section{Proposed Methodology}

\subsection{Overview}

There are two main philosophies underlying the PPO-LLM framework which are mutually reinforcing. One philosophy states that a large language model receiving as input compact and semantically clean code will be much more reliable as a generator of tests compared to the same large language model struggling to understand a convoluted and redundant program source. The second philosophy states that there is no one perfect approach for generating prompts - it all depends on the program itself, previous experience with the task, and remaining budget of generations.

\textbf{Stage I} reduces source code size while preserving logical equivalence, using a Tree-of-Thought optimization agent. \textbf{Stage II} runs an iterative test generation loop in which a PPO policy network selects, episode by episode, which of eight specialized prompt templates to apply to the optimized code, guided by a reward signal that reflects coverage improvement and branch-exploration depth.

% --- Flowchart ---
\begin{figure*}[h]
    \centering
    \includegraphics[width=0.9\linewidth]{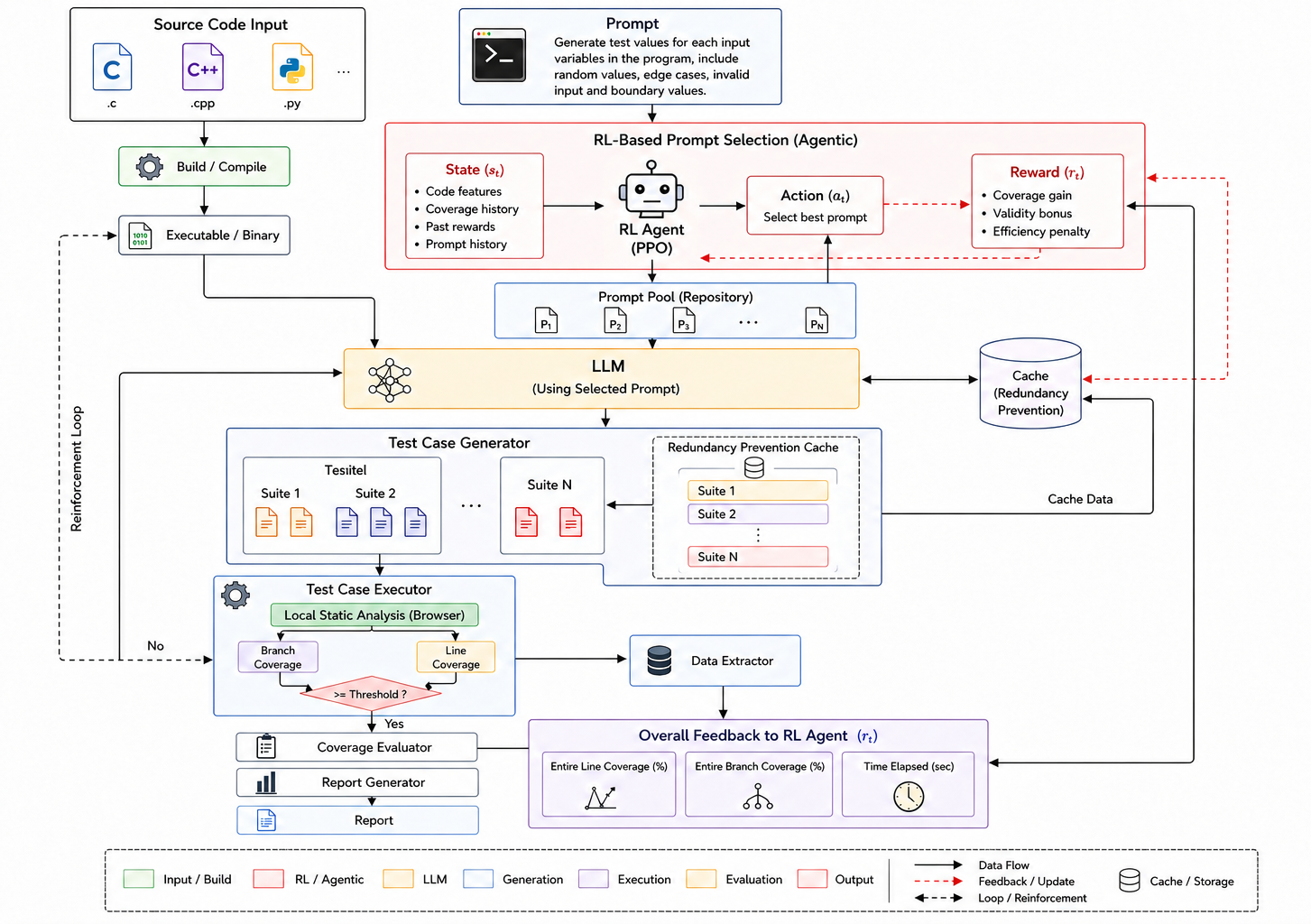}
    \caption{Overview of proposed PPO-LLM framework}
    \label{fig:Flowchart}
\end{figure*}

% --- ARCHITECTURE ---
% \begin{figure*}[h]
%     \centering
%     \includegraphics[width=0.75\linewidth]{Framework.jpeg}
% \caption{Overview of proposed PPO-LLM driven Agentic Pipeline for Adaptive Prompt Selection and Test Case Generation}
%     \label{fig:Framework}
% \end{figure*}

\subsection{Stage I: Code Optimization Component}

The optimization agent receives the raw C source file and proceeds in three steps: fragmentation, Tree-of-Thought refactoring, and equivalence verification.

\textbf{Fragmentation:} The agent first discovers semantically coherent modules, usually single functions or clusters of functions and breaks up the source code into fragments $f_1, f_2, \ldots, f_k$ at the same time storing information about the dependencies between the fragments. This helps avoid contract violations when refactoring the source code.

\textbf{Tree-of-Thought Refactoring:} Each fragment $f_i$ is submitted to an LLM optimization agent governed by the core instruction. Through its ToT representation, the model is compelled to list a number of possible refactoring options, rate them according to defined quality attributes, and then choose only the option with the highest rating. This typically results in the removal of dead code, combining repeated variable assignment, simplifying complex conditional expressions, and collapsing loops into single statements.

\textbf{Equivalence Verification:} The LLM agent works independently on the refactored fragments to verify their equivalence with the corresponding $f_i$ based on executing representative execution paths. Fragments that are found equivalent are taken as valid whereas those found to be different are resubmitted for another cycle of optimizations or left in their initial form.

The output of Stage I is the optimized code set, whose substantially reduced line count reduction in practice directly lowers the token count presented to the Stage II generation agent, reducing hallucination risk and improving semantic coherence.

\subsection{Stage II: PPO-LLM Driven Test Case Generation}
\subsubsection{MDP Formulation}
We consider test case generation to be a finite horizon Markov Decision Process $\mathcal{M} = \langle \mathcal{S}, \mathcal{A}, \mathcal{R}, \mathcal{P}, \gamma \rangle$ with the following components:

\begin{itemize}
    \item \textbf{State space} $\mathcal{S}$: A 11-dimensional continuous vector representing static code features and dynamic coverage metrics.
    \item \textbf{Action space} $\mathcal{A}$: A discrete action set containing eight indices of different prompt templates, such that $|\mathcal{A}| = 8$.
    \item \textbf{Reward function} $\mathcal{R}$: A combined reward signal that is described in Section~\ref{subsec:reward}.
    \item Transition dynamics $\mathcal{P}$: Defined based on the response of the LLM to the chosen prompt, as well as resulting coverage change.
    \item \textbf{Discount factor} $\gamma = 0.99$.
\end{itemize}

\subsubsection{State Representation}

The state $s_t \in \mathbb{R}^{11}$ observed at episode $t$ is constructed as follows:

\[\begin{aligned}
s_t = \Bigg[ &\frac{\text{LOC}}{10000},\; \frac{N_f}{50},\; \frac{N_b}{100},\; \frac{N_l}{50},\; \frac{CC}{100},\; \frac{LC_t}{100}, \\
&\frac{BC_t}{100},\; \mathbf{1}_C,\; \mathbf{1}_{Py},\; \mathbf{1}_{C{++}},\; \frac{t}{T} \Bigg]
\end{aligned}\]

where $\text{LOC}$ is the line count of the optimized code; $N_f$, $N_b$, $N_l$ are the number of functions, number of branches, and number of loops; $CC$ is the cyclomatic complexity; $LC_t$ and $BC_t$ are the cumulative percentages of line and branch coverage of the lines and branches at episode $t$; $\mathbf{1}_C$, $\mathbf{1}_{Py}$, $\mathbf{1}_{C++}$ are language indicator vectors and $t/T$ is relative generation time.

All features are clipped to $[0, 1]$ to stabilize gradient magnitudes during policy updates.

The state vector encodes code structural features and live coverage statistics. The conceptual labels "past rewards" and "prompt history" shown in Fig.~\ref{fig:Flowchart} are implicitly captured through the cumulative coverage terms $LC_t$, $BC_t$ and the relative progress indicator $t/T$ rather than as explicit dimensions.

\subsubsection{Action Space: Eight Prompt Templates}

Each of the eight templates included in $\mathcal{A}$ focuses on a different dimension of test coverage:

\begin{enumerate}
    \item \textbf{Boundary Value Analysis (BVA):} Identifies test cases that correspond to precise boundary points and beyond.
    \item \textbf{Branch Coverage Explorer (BCE):} Identifies each branching condition and produces inputs for the true and false branches.
    \item \textbf{Edge Case Hunter (ECH):} Focuses on extreme values, null inputs, and hostile input.
    \item \textbf{Error Path Exerciser (EPE):} Attempts to trigger error handling paths, division by zero conditions and malformed inputs.
    \item \textbf{Loop Boundary Tester (LBT):} Creates test inputs for zero, one, and maximum iterations in a loop.
    \item \textbf{Data Type Stress (DTS):} Tests for data type overflows, type conversion boundaries, and leading zeros in numbers.
    \item \textbf{Path Coverage Maximizer (PCM):} Constructs the control flow graph and identifies test inputs that force certain deep paths.
    \item \textbf{Random Fuzzing (FUZZ):} Creates maximal diversity in test inputs to uncover unforeseen execution paths.
\end{enumerate}

% ---------------------------------------------------------------
% PROMPTS
% ---------------------------------------------------------------
\subsubsection{Prompt Constructions for Each Template}

The prompt for each template consists of two components: an instruction component specific to each strategy, which directs the LLM What to pay attention to and a common tail component that gets appended to every call. The exact text makes a difference since it affects which paths the LLM will follow, which, in turn, affects the reward signal provided to the PPO agent. The common tail component of all eight templates is:

\begin{quote}
\ttfamily\small
Current cumulative coverage so far: Line \{$X$\}\%, Branch \{$Y$\}\%.\\
Already-generated inputs (avoid duplicates): [\ldots]\\[4pt]
RULES:\\
1.Inputs MUST be strictly printable ASCII (codes 32-126) plus newline and tab. NO null bytes or binary.\\
2.Act as a C interpreter: TRACE execution and CALCULATE EXACT stdout.\\
3.If the program prints nothing,expectedOutput = ``''.\\
4.Every test MUST be unique vs existing inputs above.\\[4pt]
Generate EXACTLY \{$N$\} test cases. Return ONLY valid JSON (no markdown fences, no prose).
\end{quote}

\noindent The strategy-specific instruction prepended to this tail for each template is given below.

\smallskip\noindent
\textbf{BVA:}
\begin{quote}
\ttfamily\small
You are an expert test engineer using Boundary Value Analysis (BVA).
STRATEGY: For every input variable, identify its valid range and generate tests at: Minimum valid value, minimum+1; Maximum valid value, maximum-1; Zero,Negative zero,Empty; Just below minimum (invalid),Just above maximum (invalid); Powers-of-2 boundaries (127,128,255,256,32767,32768,65535,65536).
\end{quote}

\smallskip\noindent
\textbf{BCE:}
\begin{quote}
\ttfamily\small
You are a Branch Coverage specialist using systematic condition analysis.
STRATEGY: Enumerate EVERY branch point (if/else,switch/case,ternary,short-circuit \&\&/||). For each branch: (1) Create input forcing the TRUE path; (2) Create input forcing the FALSE path; (3) Create input hitting the boundary condition exactly. Focus on branches NOT yet covered (current branch coverage: \{$Y$\}\%).
\end{quote}

\smallskip\noindent
\textbf{ECH:}
\begin{quote}
\ttfamily\small
You are an Edge Case specialist finding unusual failure modes.
STRATEGY: Generate extreme and unusual inputs : Very large numbers (INT\_MAX,LONG\_MAX,999999999); Very small/negative (INT\_MIN,-999999999); Empty input,single character,very long strings(100+chars); Special characters: spaces only,tabs,mixed whitespace; Repeated patterns,alternating patterns; Multiple valid inputs on separate lines vs single line.
\end{quote}

\smallskip\noindent
\textbf{EPE:}
\begin{quote}
\ttfamily\small
You are an Error Path testing specialist.
STRATEGY: Create inputs that trigger error/failure handling : Invalid format (letters when numbers expected,wrong delimiters); Missing required input fields; Division by zero conditions; Array/buffer boundary violations; Negative counts or sizes; Overflow/underflow triggers; Malformed input that reaches error branches; EOF/premature termination scenarios.
\end{quote}

\smallskip\noindent
\textbf{LBT:}
\begin{quote}
\ttfamily\small
You are a Loop Testing specialist.
STRATEGY: For each loop in the code, create inputs that: Skip the loop entirely (0 iterations); Execute exactly 1 iteration; Execute exactly 2 iterations; Execute the typical/expected number of iterations; Execute maximum possible iterations; Trigger early break/continue/return from within the loop; Test loop counter overflow or underflow; Test nested loop combinations.
\end{quote}

\smallskip\noindent
\textbf{DTS:}
\begin{quote}
\ttfamily\small
You are a Data Type Stress testing specialist.
STRATEGY: Test type-specific boundaries and conversions : Integer limits: 0,-1,1,127,-128,255,256,32767,-32768,
2147483647,-2147483648;
Floating point: 0.0,-0.0,very small(0.0001),very large(1e38); String lengths: 0,1,typical,very long; Leading zeros(007,0123); Whitespace padding before/after numbers; Scientific notation (1e5,2E-3).
\end{quote}

\smallskip\noindent
\textbf{PCM:}
\begin{quote}
\ttfamily\small
You are a Path Coverage specialist using Control Flow Analysis.
STRATEGY: Enumerate distinct execution paths from entry to exit: (1) Draw the control flow graph mentally; (2) List unique paths through the graph; (3) For each uncovered path, find the simplest input that forces that exact path; (4) Prioritise paths reaching lines NOT yet covered (current line coverage: \{$X$\}\%); (5) Target deep nesting levels and rarely reached-code sections; (6) Consider early returns and exception paths.
\end{quote}

\smallskip\noindent
\textbf{FUZZ:}
\begin{quote}
\ttfamily\small
You are a Fuzz Testing specialist generating creative, diverse inputs.
STRATEGY: Generate maximally diverse inputs using multiple strategies : Random valid inputs from different value ranges; Inputs mixing multiple data types (numbers+text); Inputs with unusual but valid formatting; Adversarial patterns that parsers might mishandle; Inputs combining multiple edge cases simultaneously; Inputs inspired by common vulnerability patterns;
Permutations and combinations of basic valid inputs; Stress tests with repeated characters or patterns. Maximise DIVERSITY - each test should be as different as possible from all others.
\end{quote}
% ---------------------------------------------------------------
% END 
% ---------------------------------------------------------------

\subsubsection{Policy Network Architecture}

The policy $\pi_\theta$ and value function $V_\phi$ share a common feature extractor. The network architecture is a two-layer MLP:
\[\mathbf{h} = \text{ReLU}\!\left(\mathbf{W}_1 s_t+\mathbf{b}_1\right),\quad \mathbf{W}_1 \in \mathbb{R}^{32 \times 11}\]
\[\pi_\theta(a \mid s_t) = \text{softmax}\!\left(\mathbf{W}_P \mathbf{h}+\mathbf{b}_P\right),\quad \mathbf{W}_P \in \mathbb{R}^{8 \times 32}\]
\[V_\phi(s_t) = \mathbf{W}_V \mathbf{h}+b_V,\quad \mathbf{W}_V \in \mathbb{R}^{1 \times 32}\]
All weight matrices are initially defined using the Xavier method. The network architecture is carefully designed to ensure that the training procedure converges within single generation run and no preliminary training process is needed.

\subsubsection{Reward Function}
\label{subsec:reward}

The reward at each episode is computed as:

\[\begin{aligned}
r_t &= 0.4 \cdot \Delta LC + 0.5 \cdot \Delta BC + 0.1 \cdot \frac{N_{uniq}}{B} \cdot 10 \\
&\quad - 0.3 \cdot \rho_{untest} + 0.1 \cdot \min\!\left(\frac{R_{red}}{100}, 0.5\right)
\end{aligned}\]

where $\Delta LC = LC_t - LC_{t-1}$ and $\Delta BC = BC_t - BC_{t-1}$ are the line and branch coverage improvements; $N_{uniq}/B$ is the fraction of batch-size $B$ tests that are genuinely unique; $\rho_{untest}$ is the ratio of still-untested branches to total branches (the penalty term) and $R_{red}$ is the percentage LOC reduction from Stage I (the optimization). The weight of branch coverage is greater than that of line coverage since it is often difficult to attain and can be used as a better indicator of faults in the code.

The reward components labelled "validity bonus" and "efficiency penalty" in Fig.~\ref{fig:Flowchart} correspond respectively to the uniqueness ratio $N_{uniq}/B$ and the untested branch penalty $\rho_{untest}$ in the equation above.

\subsubsection{PPO Update Rule}

Weights are updated for each episode using the following PPO clipped surrogate objective function:

\[\mathcal{L}^{CLIP}(\theta)=\mathbb{E}_t\!\left[\min\!\left(r_t(\theta)\hat{A}_t,\;\text{clip}\!\left(r_t(\theta),1-\epsilon,1+\epsilon\right)\hat{A}_t\right)\right]\]

where $r_t(\theta) = \pi_\theta(a_t \mid s_t) / \pi_{\theta_{old}}(a_t \mid s_t)$ represents probability ratio, $\hat{A}_t$ is the Generalized Advantage Estimation (GAE) value with $\lambda = 0.95$, and $\epsilon = 0.2$ is the clipping threshold. An entropy bonus with coefficient $0.02$ is added to the objective to prevent premature policy collapse. The full loss is:

\[\mathcal{L}(\theta)=-\mathcal{L}^{CLIP}(\theta)+0.5 \cdot \mathcal{L}^{VF}-0.02 \cdot \mathcal{H}[\pi_\theta]\]

where $\mathcal{L}^{VF} = \mathbb{E}_t[(V_\phi(s_t) - \hat{R}_t)^2]$ represents value function loss and $\mathcal{H}[\pi_\theta]$ is the policy entropy. Gradient updates use stochastic gradient descent with gradient norm clipping at 1.0.

\subsection{PPO-LLM Algorithm}

\begin{algorithm}
\caption{PPO-LLM Test Case Generation Procedure}
\label{alg:ppori}
\begin{algorithmic}[1]
\REQUIRE Dataset $D$, code parameters, generation budget $T$, batch size $B$
\ENSURE High-coverage test suite $\mathcal{T}$

\textit{// Stage-I: Code Optimization}
\STATE $F \leftarrow \texttt{Fragmentize}(D)$
\FORALL{$f \in F$}
    \STATE $f_{opt} \leftarrow \texttt{ToT\_Optimize}(f)$
    \STATE $\texttt{Verify}(f_{opt} \equiv f)$
\ENDFOR
\STATE $D_{opt} \leftarrow \texttt{Assemble}(F_{opt})$

\textit{// Stage-II: PPO-LLM Guided Test Case Generation}
\STATE Initialize PPO network $\pi_\theta$, value network $V_\phi$
\STATE $\mathcal{T} \leftarrow \emptyset$;\ $\mathcal{K} \leftarrow \emptyset$ (uniqueness cache)
\FOR{episode $t = 1$ to $T$}
    \STATE $s_t \leftarrow \texttt{GetState}(\textit{code features}, LC_t, BC_t, t)$
    \STATE $a_t \sim \pi_\theta(\cdot \mid s_t)$ \quad \textit{// sample template}
    \STATE $p_t \leftarrow \texttt{Template}[a_t]\texttt{.buildPrompt}(D_{opt}, \mathcal{K}, B,$ \\ 
       \ \ \ \ \ $LC_t, BC_t)$
    \STATE $\tau \leftarrow \texttt{LLM\_Generate}(p_t)$
    \STATE $\tau^* \leftarrow \texttt{FilterUnique}(\tau, \mathcal{K})$;\ $\mathcal{K} \leftarrow \mathcal{K} \cup \tau^*$
    \STATE $(LC_{t+1}, BC_{t+1}) \leftarrow \texttt{EvalCoverage}(\tau^*)$
    \STATE $r_t \leftarrow \texttt{ComputeReward}(LC_t, LC_{t+1}, BC_t, BC_{t+1}, \ldots)$
    \STATE $\texttt{StoreTransition}(s_t, a_t, r_t, s_{t+1})$
    \STATE $\texttt{PPO\_Update}(\pi_\theta, V_\phi)$
    \STATE $\mathcal{T} \leftarrow \mathcal{T} \cup \tau^*$
\ENDFOR
\RETURN $\mathcal{T}$
\end{algorithmic}
\end{algorithm}

% ============================================================
\section{Dataset Description and Metrics}

\subsection{Benchmark Programs}

PPO-LLM is assessed on 20 benchmark programs taken from well-known benchmark sets PALS (Process Algebraic Labeled Systems) and RERS (Rigorous Examination of Reactive Systems). These benchmarks constitute industrial-level C code snippets that vary significantly in terms of their sizes (291 to 9,111 LOCs), structural complexities and loop-bound depths. Each benchmark undergoes testing under eight different loop-bound settings: BOUND 1, 10, 100, 250, 500, 1000, 1500 and 2000, resulting in 160 testing instances. The value of step parameter ($k$) is set to 10 for all tests. In addition, any instance that takes more than one hour to solve using CBMC is deemed to have been timed out.

\subsection{Metrics}

\subsubsection{Branch Coverage}

Branch coverage is defined to be the ratio of number of conditional branches that are executed from the program in a set of test cases used to test the program. If there are $N_b$ total branches in a program and $B_{cov}$ are the branches covered, then
\[\text{Branch Coverage} = \frac{B_{cov}}{N_b} \times 100\%\]
Branch coverage has been described as one of the most challenging metrics for structural testing, since it calls for every node in the control flow diagram, which represents every decision (like if, else, switch cases, loop conditions, and short circuits), to be executed once in both their true and false forms.

\subsubsection{Line Coverage}

Line coverage measures the percentage of the executable lines of code that have been exercised during testing:
\[\text{Line Coverage} = \frac{L_{cov}}{L_{total}} \times 100\% \]
where $L_{total}$ represents total number of executable lines and $L_{cov}$ denotes those lines that have been executed using at least one test case. Line coverage is easier to meet than branch coverage since there is no need to cover all alternative paths to achieve line coverage.

\subsubsection{LOC Reduction}

In order to measure the success of Stage I, we will calculate:
\[\text{LOC Reduction (\%)} = \frac{\text{LOC}_{orig} - \text{LOC}_{opt}}{\text{LOC}_{orig}} \times 100\% \]
If there is a high percentage of reduction, this implies that the optimization tool was able to remove redundant LOCs. This measure is important as it is directly linked to the load faced by the LLM in Stage II, as more focused code fragment allows the model to result in better quality test cases with fewer hallucinations.

% ============================================================
\section{Results and Discussion}

The branch and line coverage rates obtained by CBMC, kS-LLM, kS-LLM++ and PPO-LLM are provided in Tables~\ref{tab:coverage_1} 
and~\ref{tab:coverage_2} on 20 benchmarking problems across 8 bound values. The numbers speak for themselves and clearly show that PPO-LLM outperforms other approaches in the majority of experiments.

% =========================================================
% TABLE 1: Bounds 1, 10, 100, 250
% =========================================================
\begin{table}[htbp]
\caption{Branch and Line Coverage Analysis (Bounds 1 -- 250)}
\label{tab:coverage_1}
\centering
\footnotesize 
\renewcommand{\arraystretch}{1.05} 
\setlength{\tabcolsep}{2pt}
\resizebox{\columnwidth}{!}{%
\begin{tabular}{|l|c|r|r|>{}r|r|r|r|>{}r|r|}
\hline
\rowcolor{white}
\textbf{Program Name} & \textbf{LOC} & \multicolumn{4}{c|}{\textbf{Branch Coverage (\%)}} & \multicolumn{4}{c|}{\textbf{Line Coverage (\%)}} \\
\cline{3-10}
\rowcolor{white}
& & \textbf{CBMC} & \textbf{kS-LLM} & \textbf{kS-LLM++} & \textbf{PPO} & \textbf{CBMC} & \textbf{kS-LLM} & \textbf{kS-LLM++} & \textbf{PPO} \\
\hline
\multicolumn{10}{|c|}{\cellcolor{orange!80}\textbf{BOUND 1}} \\
\hline
Mpals1-B10-cil.c & 295 & 16.25 & 23.75 & 73.4 & \cellcolor{green}100 & 48.23 & 57.45 & 90.2 & \cellcolor{green}91.2 \\ \hline
Mpals2-B10-cil.c & 370 & 12.04 & 20.37 & 62.8 & \cellcolor{green}100 & 43.95 & 47.77 & 76.1 & \cellcolor{green}90.8 \\ \hline
Mpals3-B10-cil.c & 517 & 1.66 & 1.79 & 44.4 & \cellcolor{green}98.8 & 10.57 & 10.57 & 55.8 & \cellcolor{green}91.8 \\ \hline
Mpals23-B10-cil.c & 1624 & 12.18 & 26.42 & 62.8 & \cellcolor{green}100 & 23.56 & 27.24 & 77.6 & \cellcolor{green}92.6 \\ \hline
Mtest20-B10-cil.c & 348 & 12.04 & 20.37 & 60.6 & \cellcolor{green}100 & 40.72 & 46.11 & 77.2 & \cellcolor{green}90.8 \\ \hline
pals2-B10-cil.c & 370 & 12.04 & 20.37 & 71.7 & \cellcolor{green}97.4 & 44.23 & 48.08 & 86.6 & \cellcolor{green}90.9 \\ \hline
pals3-B10-cil.c & 6254 & 1.43 & 1.43 & 62.9 & \cellcolor{green}98.7 & 2.38 & 2.38 & 78.3 & \cellcolor{green}91.6 \\ \hline
pals22-B10-cil.c & 764 & 1.23 & 13.44 & 67.1 & \cellcolor{green}96.2 & 6.93 & 20.44 & 84.0 & \cellcolor{green}88.9 \\ \hline
pals23-B10-cil.c & 291 & 16.25 & 21.25 & 73.6 & \cellcolor{green}96.2 & 48.57 & 52.14 & 87.7 & \cellcolor{green}88.4 \\ \hline
PS-P1-L-R18-B4.c & 295 & 16.25 & 16.25 & 45.3 & \cellcolor{red}6.9 & 48.89 & 48.89 & 57.3 & \cellcolor{red}25.6 \\ \hline
PS-P1-L-T-R16-B2.c & 1576 & 15.41 & 17.86 & 82.7 & \cellcolor{red}4.9 & 27.58 & 27.58 & 93.6 & \cellcolor{red}21.4 \\ \hline
PS-P1-L-T-R20-B1.c & 998 & 2.16 & 15.57 & 44.3 & \cellcolor{red}1.7 & 6.59 & 14.73 & 55.5 & \cellcolor{red}13 \\ \hline
PS-P1-NT-R14-B4.c & 291 & 16.25 & 16.25 & 86.8 & \cellcolor{red}31.9 & 48.57 & 48.57 & 98.6 & \cellcolor{red}12.2 \\ \hline
PS-P1-WB-R15-B5.c & 296 & 16.25 & 16.25 & 43.6 & \cellcolor{green}47.4 & 48.89 & 48.89 & 54.8 & \cellcolor{red}37.9 \\ \hline
PS-P1-WB-T-R15-B1.c & 2395 & 11.01 & 17.85 & 69.9 & \cellcolor{red}37.8 & 17.09 & 21.18 & 82.1 & \cellcolor{red}32.9 \\ \hline
PS-P2-A-R14-B6.c & 9111 & 0.38 & 2.15 & 43.4 & \cellcolor{red}38 & 0.90 & 2.66 & 56.7 & \cellcolor{red}12.1 \\ \hline
PS-P2-L-R16-B3.c & 293 & 16.25 & 16.25 & 43.5 & \cellcolor{green}52.1 & 48.23 & 48.23 & 55.9 & \cellcolor{green}85.4 \\ \hline
PS-P2-L-R18-B7.c & 667 & 3.68 & 46.10 & 49.5 & \cellcolor{red}3.1 & 7.98 & 62.97 & 64.3 & \cellcolor{red}35.4 \\ \hline
PS-P4-L-T-R20-B1.c & 2285 & 11.34 & 12.42 & 44.0 & \cellcolor{red}0.8 & 18.24 & 18.24 & 55.4 & \cellcolor{red}7.9 \\ \hline
PS-Prob1-IO-R14-B7.c & 1441 & 11.19 & 12.58 & 44.9 & \cellcolor{red}3.3 & 25.94 & 25.94 & 58.6 & \cellcolor{red}14.4 \\ \hline
\multicolumn{10}{|c|}{\cellcolor{orange!80}\textbf{BOUND 10}} \\
\hline
Mpals1-B10-cil.c & 295 & 31.25 & 40.0 & 45.5 & \cellcolor{green}100 & 63.83 & 66.67 & 59.0 & \cellcolor{green}90.2 \\ \hline
Mpals2-B10-cil.c & 370 & 17.59 & 31.48 & 43.4 & \cellcolor{green}100 & 52.23 & 59.87 & 54.8 & \cellcolor{green}90.8 \\ \hline
Mpals3-B10-cil.c & 517 & 1.66 & 2.17 & 44.3 & \cellcolor{green}100 & 10.57 & 10.57 & 57.0 & \cellcolor{green}91.8 \\ \hline
Mpals23-B10-cil.c & 1624 & 12.18 & 15.01 & 45.4 & \cellcolor{green}100 & 23.56 & 23.56 & 55.5 & \cellcolor{green}92.6 \\ \hline
Mtest20-B10-cil.c & 348 & 17.59 & 31.48 & 45.0 & \cellcolor{green}100 & 48.50 & 57.49 & 56.5 & \cellcolor{green}90.8 \\ \hline
pals2-B10-cil.c & 370 & 17.59 & 34.26 & 45.0 & \cellcolor{green}96.1 & 52.56 & 62.18 & 55.9 & \cellcolor{green}90.9 \\ \hline
pals3-B10-cil.c & 6254 & 66.94 & 10.14 & 44.5 & \cellcolor{green}96.2 & 100.0 & 15.0 & 57.8 & \cellcolor{green}84.8 \\ \hline
pals22-B10-cil.c & 764 & 1.23 & 30.83 & 44.9 & \cellcolor{green}95.2 & 6.93 & 44.89 & 56.4 & \cellcolor{green}79.5 \\ \hline
pals23-B10-cil.c & 291 & 32.5 & 43.75 & 44.5 & \cellcolor{green}95.2 & 64.29 & 70.0 & 57.7 & \cellcolor{green}79.5 \\ \hline
PS-P1-L-R18-B4.c & 295 & 32.5 & 38.75 & 45.7 & \cellcolor{red}6.9 & 65.19 & 67.41 & 58.3 & \cellcolor{red}25.6 \\ \hline
PS-P1-L-T-R16-B2.c & 1576 & 50.0 & 20.20 & 45.2 & \cellcolor{red}5.9 & 83.53 & 31.15 & 57.7 & \cellcolor{red}21.4 \\ \hline
PS-P1-L-T-R20-B1.c & 998 & 2.16 & 10.69 & 87.6 & \cellcolor{red}3.2 & 6.59 & 11.36 & 98.6 & \cellcolor{red}13.3 \\ \hline
PS-P1-NT-R14-B4.c & 291 & 32.5 & 38.75 & 56.8 & \cellcolor{red}30.4 & 64.29 & 66.43 & 69.0 & \cellcolor{red}12 \\ \hline
PS-P1-WB-R15-B5.c & 296 & 32.5 & 38.75 & 39.7 & \cellcolor{green}47 & 65.19 & 67.41 & 50.8 & \cellcolor{red}37.8 \\ \hline
PS-P1-WB-T-R15-B1.c & 2395 & 11.01 & 17.94 & 64.1 & \cellcolor{red}37.5 & 17.09 & 21.51 & 76.5 & \cellcolor{red}32.8 \\ \hline
PS-P2-A-R14-B6.c & 9111 & 0.38 & 5.07 & 73.0 & \cellcolor{red}42.8 & 0.90 & 6.31 & 83.5 & \cellcolor{red}14.5 \\ \hline
PS-P2-L-R16-B3.c & 293 & 32.5 & 38.75 & 43.4 & \cellcolor{red}4.1 & 63.83 & 66.67 & 56.0 & \cellcolor{red}28.9 \\ \hline
PS-P2-L-R18-B7.c & 667 & 3.68 & 41.99 & 35.6 & \cellcolor{red}3.1 & 7.98 & 61.42 & 48.9 & \cellcolor{red}30.3 \\ \hline
PS-P4-L-T-R20-B1.c & 2285 & 45.03 & 13.99 & 80.9 & \cellcolor{red}0.3 & 73.09 & 19.96 & 92.2 & \cellcolor{red}7.7 \\ \hline
PS-Prob1-IO-R14-B7.c & 1441 & 21.42 & 21.25 & 54.0 & \cellcolor{red}3.3 & 42.70 & 37.39 & 65.6 & \cellcolor{red}15.1 \\ \hline
\multicolumn{10}{|c|}{\cellcolor{orange!80}\textbf{BOUND 100}} \\
\hline
Mpals1-B10-cil.c & 295 & 31.25 & 41.25 & 64.8 & \cellcolor{green}98.8 & 63.83 & 66.67 & 81.7 & \cellcolor{green}91.2 \\ \hline
Mpals2-B10-cil.c & 370 & 17.59 & 31.48 & 74.2 & \cellcolor{green}100 & 52.23 & 59.87 & 87.2 & \cellcolor{green}90.8 \\ \hline
Mpals3-B10-cil.c & 517 & 1.66 & 2.17 & 74.2 & \cellcolor{green}98.8 & 10.57 & 10.57 & 90.0 & \cellcolor{green}91.8 \\ \hline
Mpals23-B10-cil.c & 1624 & 12.18 & 29.76 & 53.1 & \cellcolor{green}100 & 23.56 & 35.16 & 66.8 & \cellcolor{green}92.6 \\ \hline
Mtest20-B10-cil.c & 348 & 17.59 & 31.48 & 72.4 & \cellcolor{green}100 & 48.50 & 57.49 & 85.5 & \cellcolor{green}90.8 \\ \hline
pals2-B10-cil.c & 370 & 17.59 & 31.48 & 70.8 & \cellcolor{green}97.4 & 52.56 & 60.26 & 87.3 & \cellcolor{green}90.9 \\ \hline
pals3-B10-cil.c & 6254 & 0.0 & 10.14 & 68.5 & \cellcolor{green}96.2 & 0.0 & 15.0 & 84.4 & \cellcolor{green}84.8 \\ \hline
pals22-B10-cil.c & 764 & 1.23 & 29.88 & 69.5 & \cellcolor{green}97.1 & 6.93 & 41.24 & 85.6 & \cellcolor{red}79.5 \\ \hline
pals23-B10-cil.c & 291 & 32.5 & 38.75 & 61.5 & \cellcolor{green}96.2 & 64.29 & 66.43 & 77.1 & \cellcolor{green}88.4 \\ \hline
PS-P1-L-R18-B4.c & 295 & 32.5 & 38.75 & 55.9 & \cellcolor{green}100 & 65.19 & 67.41 & 69.3 & \cellcolor{green}82.6 \\ \hline
PS-P1-L-T-R16-B2.c & 1576 & 50.0 & 20.82 & 65.2 & \cellcolor{red}52.9 & 83.53 & 31.15 & 78.3 & \cellcolor{green}82 \\ \hline
PS-P1-L-T-R20-B1.c & 998 & 2.16 & 12.20 & 50.8 & \cellcolor{red}0.8 & 6.59 & 12.06 & 65.3 & \cellcolor{red}12.8 \\ \hline
PS-P1-NT-R14-B4.c & 291 & 32.5 & 38.75 & 57.0 & \cellcolor{red}28.6 & 64.29 & 66.43 & 71.6 & \cellcolor{red}12 \\ \hline
PS-P1-WB-R15-B5.c & 296 & 32.5 & 38.75 & 33.2 & \cellcolor{green}47.2 & 65.19 & 67.41 & 44.3 & \cellcolor{red}38 \\ \hline
PS-P1-WB-T-R15-B1.c & 2395 & 11.01 & 17.24 & 53.7 & \cellcolor{red}37.5 & 17.09 & 20.04 & 67.3 & \cellcolor{red}32.8 \\ \hline
PS-P2-A-R14-B6.c & 9111 & 0.38 & 14.22 & 51.4 & \cellcolor{red}41 & 0.90 & 17.30 & 65.7 & \cellcolor{red}14.3 \\ \hline
PS-P2-L-R16-B3.c & 293 & 32.5 & 38.75 & 62.9 & \cellcolor{red}52.1 & 63.83 & 66.67 & 74.7 & \cellcolor{green}85.4 \\ \hline
PS-P2-L-R18-B7.c & 667 & 3.68 & 44.37 & 47.7 & \cellcolor{red}3.1 & 7.98 & 65.53 & 64.1 & \cellcolor{red}30.8 \\ \hline
PS-P4-L-T-R20-B1.c & 2285 & 0.0 & 15.31 & 45.6 & \cellcolor{red}0.5 & 0.0 & 19.96 & 60.4 & \cellcolor{red}7.3 \\ \hline
PS-Prob1-IO-R14-B7.c & 1441 & 21.53 & 21.25 & 63.3 & \cellcolor{red}3.3 & 42.70 & 37.39 & 75.9 & \cellcolor{red}15.1 \\ \hline
\multicolumn{10}{|c|}{\cellcolor{orange!80}\textbf{BOUND 250}} \\
\hline
Mpals1-B10-cil.c & 295 & 31.25 & 40.0 & 48.8 & \cellcolor{green}98.8 & 63.83 & 66.67 & 59.8 & \cellcolor{green}91.2 \\ \hline
Mpals2-B10-cil.c & 370 & 17.59 & 31.48 & 47.5 & \cellcolor{green}100 & 52.23 & 59.87 & 59.2 & \cellcolor{green}90.8 \\ \hline
Mpals3-B10-cil.c & 517 & 1.66 & 2.69 & 42.7 & \cellcolor{green}98.8 & 10.57 & 10.57 & 55.0 & \cellcolor{green}91.8 \\ \hline
Mpals23-B10-cil.c & 1624 & 12.18 & 28.13 & 49.3 & \cellcolor{green}100 & 23.56 & 33.93 & 61.8 & \cellcolor{green}92.6 \\ \hline
Mtest20-B10-cil.c & 348 & 17.59 & 31.48 & 45.9 & \cellcolor{green}100 & 48.50 & 57.49 & 58.8 & \cellcolor{green}90.8 \\ \hline
pals2-B10-cil.c & 370 & 17.59 & 31.48 & 72.9 & \cellcolor{green}96.1 & 52.56 & 60.26 & 85.4 & \cellcolor{green}90.9 \\ \hline
pals3-B10-cil.c & 6254 & 0.0 & 10.14 & 43.4 & \cellcolor{green}96.2 & 0.0 & 15.0 & 54.9 & \cellcolor{green}84.8 \\ \hline
pals22-B10-cil.c & 764 & 1.23 & 8.59 & 52.7 & \cellcolor{green}96.2 & 6.93 & 11.68 & 66.0 & \cellcolor{green}88.9 \\ \hline
pals23-B10-cil.c & 291 & 32.5 & 38.75 & 72.4 & \cellcolor{green}96.2 & 64.29 & 66.43 & 84.8 & \cellcolor{green}88.4 \\ \hline
PS-P1-L-R18-B4.c & 295 & 32.5 & 43.75 & 73.5 & \cellcolor{green}100 & 65.19 & 71.11 & 84.0 & \cellcolor{red}82.6 \\ \hline
PS-P1-L-T-R16-B2.c & 1576 & 0.0 & 20.0 & 33.1 & \cellcolor{red}4.9 & 0.0 & 31.15 & 45.7 & \cellcolor{red}18.9 \\ \hline
PS-P1-L-T-R20-B1.c & 998 & 2.16 & 11.82 & 63.0 & \cellcolor{red}0.8 & 6.59 & 11.36 & 75.1 & \cellcolor{red}12.7 \\ \hline
PS-P1-NT-R14-B4.c & 291 & 32.5 & 38.75 & 53.2 & \cellcolor{red}28.3 & 64.29 & 66.43 & 65.9 & \cellcolor{red}12 \\ \hline
PS-P1-WB-R15-B5.c & 296 & 32.5 & 38.75 & 61.3 & \cellcolor{red}45 & 52.56 & 67.41 & 73.7 & \cellcolor{red}36.2 \\ \hline
PS-P1-WB-T-R15-B1.c & 2395 & 11.01 & 16.90 & 52.7 & \cellcolor{red}37.5 & 17.09 & 21.51 & 64.6 & \cellcolor{red}32.8 \\ \hline
PS-P2-A-R14-B6.c & 9111 & 0.38 & 5.12 & 69.1 & \cellcolor{red}36.8 & 0.90 & 6.36 & 81.6 & \cellcolor{red}12.2 \\ \hline
PS-P2-L-R16-B3.c & 293 & 32.5 & 38.75 & 45.8 & \cellcolor{red}4.1 & 63.83 & 66.67 & 58.0 & \cellcolor{red}31.5 \\ \hline
PS-P2-L-R18-B7.c & 667 & 3.68 & 30.09 & 42.8 & \cellcolor{red}3.1 & 7.98 & 42.57 & 55.4 & \cellcolor{red}30.9 \\ \hline
PS-P4-L-T-R20-B1.c & 2285 & 0.0 & 14.98 & 56.7 & \cellcolor{red}0.4 & 0.0 & 19.96 & 68.4 & \cellcolor{red}7.7 \\ \hline
PS-Prob1-IO-R14-B7.c & 1441 & 21.53 & 21.53 & 63.0 & \cellcolor{red}3.3 & 42.70 & 37.39 & 74.7 & \cellcolor{red}15.1 \\ \hline
\end{tabular}%
}
\end{table}

% =========================================================
% TABLE 2: Bounds 500, 1000, 1500, 2000
% =========================================================
\begin{table}[htbp]
\caption{Branch and Line Coverage Analysis (Bounds 500 -- 2000)}
\label{tab:coverage_2}
\centering
\footnotesize 
\renewcommand{\arraystretch}{1.05} 
\setlength{\tabcolsep}{2pt}
\resizebox{\columnwidth}{!}{%
\begin{tabular}{|l|c|r|r|>{}r|r|r|r|>{}r|r|}
\hline
\rowcolor{white}
\textbf{Program Name} & \textbf{LOC} & \multicolumn{4}{c|}{\textbf{Branch Coverage (\%)}} & \multicolumn{4}{c|}{\textbf{Line Coverage (\%)}} \\
\cline{3-10}
\rowcolor{white}
& & \textbf{CBMC} & \textbf{kS-LLM} & \textbf{kS-LLM++} & \textbf{PPO} & \textbf{CBMC} & \textbf{kS-LLM} & \textbf{kS-LLM++} & \textbf{PPO} \\
\hline
\multicolumn{10}{|c|}{\cellcolor{orange!80}\textbf{BOUND 500}} \\
\hline
Mpals1-B10-cil.c & 295 & 31.25 & 40.0 & 69.8 & \cellcolor{green}100 & 63.83 & 66.67 & 85.5 & \cellcolor{green}91.2 \\ \hline
Mpals2-B10-cil.c & 370 & 17.59 & 31.48 & 73.3 & \cellcolor{green}100 & 52.23 & 59.87 & 89.4 & \cellcolor{green}90.8 \\ \hline
Mpals3-B10-cil.c & 517 & 1.66 & 2.69 & 74.7 & \cellcolor{green}100 & 10.57 & 10.57 & 90.0 & \cellcolor{green}91.8 \\ \hline
Mpals23-B10-cil.c & 1624 & 12.18 & 23.93 & 68.4 & \cellcolor{green}100 & 23.56 & 26.48 & 83.0 & \cellcolor{green}92.6 \\ \hline
Mtest20-B10-cil.c & 348 & 17.59 & 31.48 & 75.6 & \cellcolor{green}100 & 48.50 & 57.49 & 91.5 & \cellcolor{red}90.8 \\ \hline
pals2-B10-cil.c & 370 & 17.59 & 31.48 & 57.9 & \cellcolor{green}97.4 & 52.56 & 60.26 & 72.9 & \cellcolor{green}90.9 \\ \hline
pals3-B10-cil.c & 6254 & 0.0 & 10.14 & 68.0 & \cellcolor{green}97.5 & 0.0 & 15.0 & 83.8 & \cellcolor{green}91.6 \\ \hline
pals22-B10-cil.c & 764 & 1.23 & 23.67 & 66.8 & \cellcolor{green}96.2 & 6.93 & 38.69 & 82.1 & \cellcolor{green}88.9 \\ \hline
pals23-B10-cil.c & 291 & 32.5 & 38.75 & 60.7 & \cellcolor{green}96.2 & 64.29 & 66.43 & 73.8 & \cellcolor{green}88.4 \\ \hline
PS-P1-L-R18-B4.c & 295 & 32.5 & 38.75 & 54.3 & \cellcolor{green}100 & 65.19 & 67.41 & 69.5 & \cellcolor{green}82.6 \\ \hline
PS-P1-L-T-R16-B2.c & 1576 & 0.0 & 21.02 & 67.2 & \cellcolor{red}4.9 & 0.0 & 31.15 & 80.8 & \cellcolor{red}24.5 \\ \hline
PS-P1-L-T-R20-B1.c & 998 & 2.16 & 18.86 & 43.0 & \cellcolor{red}2.8 & 6.59 & 8.98 & 55.1 & \cellcolor{red}13.4 \\ \hline
PS-P1-NT-R14-B4.c & 291 & 32.5 & 38.75 & 50.1 & \cellcolor{red}28.6 & 64.29 & 66.43 & 63.6 & \cellcolor{red}12 \\ \hline
PS-P1-WB-R15-B5.c & 296 & 32.5 & 38.75 & 44.1 & \cellcolor{green}46.9 & 65.19 & 67.41 & 61.0 & \cellcolor{red}37.8 \\ \hline
PS-P1-WB-T-R15-B1.c & 2395 & 11.01 & 18.02 & 49.8 & \cellcolor{red}37.8 & 17.09 & 21.92 & 65.6 & \cellcolor{red}32.7 \\ \hline
PS-P2-A-R14-B6.c & 9111 & 0.38 & 10.70 & 43.6 & \cellcolor{red}41 & 0.90 & 6.36 & 55.4 & \cellcolor{red}14 \\ \hline
PS-P2-L-R16-B3.c & 293 & 32.5 & 38.75 & 49.8 & \cellcolor{green}52.1 & 63.83 & 66.67 & 65.6 & \cellcolor{green}85.4 \\ \hline
PS-P2-L-R18-B7.c & 667 & 3.68 & 52.60 & 54.1 & \cellcolor{red}3.1 & 7.98 & 71.40 & 67.6 & \cellcolor{red}26.3 \\ \hline
PS-P4-L-T-R20-B1.c & 2285 & 0.0 & 14.65 & 69.9 & \cellcolor{red}0.4 & 0.0 & 19.96 & 82.6 & \cellcolor{red}7.7 \\ \hline
PS-Prob1-IO-R14-B7.c & 1441 & 0.0 & 21.03 & 55.0 & \cellcolor{red}3.3 & 0.0 & 37.39 & 70.0 & \cellcolor{red}14.7 \\ \hline
\multicolumn{10}{|c|}{\cellcolor{orange!80}\textbf{BOUND 1000}} \\
\hline
Mpals1-B10-cil.c & 295 & 31.25 & 40.0 & 63.5 & \cellcolor{green}100 & 63.83 & 66.67 & 79.7 & \cellcolor{green}91.2 \\ \hline
Mpals2-B10-cil.c & 370 & 17.59 & 31.48 & 73.5 & \cellcolor{green}100 & 52.23 & 59.87 & 89.0 & \cellcolor{green}90.8 \\ \hline
Mpals3-B10-cil.c & 517 & 1.66 & 2.69 & 66.2 & \cellcolor{green}100 & 10.57 & 10.57 & 80.0 & \cellcolor{green}91.8 \\ \hline
Mpals23-B10-cil.c & 1624 & 0.0 & 30.1 & 69.9 & \cellcolor{green}100 & 0.0 & 37.13 & 83.6 & \cellcolor{green}92.6 \\ \hline
Mtest20-B10-cil.c & 348 & 17.59 & 31.48 & 61.8 & \cellcolor{green}100 & 48.50 & 57.49 & 76.6 & \cellcolor{green}90.8 \\ \hline
pals2-B10-cil.c & 370 & 17.59 & 31.48 & 56.3 & \cellcolor{green}97.4 & 52.56 & 60.26 & 70.1 & \cellcolor{green}90.9 \\ \hline
pals3-B10-cil.c & 6254 & 0.0 & 10.14 & 75.7 & \cellcolor{green}97.5 & 0.0 & 15.0 & 88.8 & \cellcolor{green}91.6 \\ \hline
pals22-B10-cil.c & 764 & 1.23 & 27.76 & 59.7 & \cellcolor{green}95.2 & 6.93 & 38.69 & 76.0 & \cellcolor{green}79.5 \\ \hline
pals23-B10-cil.c & 291 & 32.5 & 38.75 & 62.5 & \cellcolor{green}96.2 & 64.29 & 66.43 & 77.6 & \cellcolor{green}88.4 \\ \hline
PS-P1-L-R18-B4.c & 295 & 32.5 & 38.75 & 54.5 & \cellcolor{green}100 & 65.19 & 67.41 & 67.6 & \cellcolor{green}82.6 \\ \hline
PS-P1-L-T-R16-B2.c & 1576 & 0.0 & 21.02 & 66.4 & \cellcolor{red}4.9 & 0.0 & 31.15 & 79.8 & \cellcolor{red}24.5 \\ \hline
PS-P1-L-T-R20-B1.c & 998 & 2.16 & 9.57 & 46.0 & \cellcolor{red}1.7 & 6.59 & 8.98 & 57.1 & \cellcolor{red}12.9 \\ \hline
PS-P1-NT-R14-B4.c & 291 & 32.5 & 38.75 & 81.5 & \cellcolor{red}28.1 & 64.29 & 66.43 & 93.6 & \cellcolor{red}11.9 \\ \hline
PS-P1-WB-R15-B5.c & 296 & 32.5 & 38.75 & 75.9 & \cellcolor{red}46.7 & 65.19 & 67.41 & 87.3 & \cellcolor{red}37.8 \\ \hline
PS-P1-WB-T-R15-B1.c & 2395 & 0.0 & 18.72 & 57.3 & \cellcolor{red}36.8 & 0.0 & 21.92 & 69.6 & \cellcolor{red}32.5 \\ \hline
PS-P2-A-R14-B6.c & 9111 & 0.38 & 5.12 & 58.4 & \cellcolor{red}37.8 & 0.90 & 6.36 & 70.8 & \cellcolor{red}13 \\ \hline
PS-P2-L-R16-B3.c & 293 & 32.5 & 38.75 & 66.8 & \cellcolor{red}4.1 & 63.83 & 66.67 & 78.0 & \cellcolor{red}31.5 \\ \hline
PS-P2-L-R18-B7.c & 667 & 3.68 & 50.43 & 69.1 & \cellcolor{red}3.1 & 7.98 & 71.40 & 82.0 & \cellcolor{red}29.8 \\ \hline
PS-P4-L-T-R20-B1.c & 2285 & 0.0 & 14.16 & 43.6 & \cellcolor{red}0.4 & 0.0 & 19.96 & 55.5 & \cellcolor{red}7.6 \\ \hline
PS-Prob1-IO-R14-B7.c & 1441 & 0.0 & 21.36 & 77.7 & \cellcolor{red}3.3 & 0.0 & 37.39 & 90.7 & \cellcolor{red}15.1 \\ \hline
\multicolumn{10}{|c|}{\cellcolor{orange!80}\textbf{BOUND 1500}} \\
\hline
Mpals1-B10-cil.c & 295 & 31.25 & 40.0 & 76.7 & \cellcolor{green}100 & 63.83 & 66.67 & 92.3 & \cellcolor{red}91.2 \\ \hline
Mpals2-B10-cil.c & 370 & 17.59 & 31.48 & 73.0 & \cellcolor{green}100 & 52.23 & 59.87 & 88.4 & \cellcolor{green}90.8 \\ \hline
Mpals3-B10-cil.c & 517 & 1.66 & 8.95 & 73.7 & \cellcolor{green}98.8 & 10.57 & 13.71 & 88.5 & \cellcolor{green}91.8 \\ \hline
Mpals23-B10-cil.c & 1624 & 0.0 & 29.76 & 64.2 & \cellcolor{green}99.1 & 0.0 & 36.19 & 77.3 & \cellcolor{green}92.6 \\ \hline
Mtest20-B10-cil.c & 348 & 17.59 & 31.48 & 74.1 & \cellcolor{green}100 & 48.50 & 57.49 & 90.4 & \cellcolor{green}90.8 \\ \hline
pals2-B10-cil.c & 370 & 17.59 & 31.48 & 71.7 & \cellcolor{green}97.4 & 52.56 & 60.26 & 84.8 & \cellcolor{green}90.9 \\ \hline
pals3-B10-cil.c & 6254 & 0.0 & 10.14 & 80.9 & \cellcolor{green}96.2 & 0.0 & 15.0 & 94.2 & \cellcolor{red}84.8 \\ \hline
pals22-B10-cil.c & 764 & 1.23 & 26.53 & 47.5 & \cellcolor{green}96.2 & 6.93 & 32.48 & 61.3 & \cellcolor{green}88.9 \\ \hline
pals23-B10-cil.c & 291 & 32.5 & 38.75 & 63.5 & \cellcolor{green}98.1 & 64.29 & 66.43 & 77.6 & \cellcolor{green}88.4 \\ \hline
PS-P1-L-R18-B4.c & 295 & 32.5 & 38.75 & 56.6 & \cellcolor{red}6.9 & 65.19 & 67.41 & 72.7 & \cellcolor{red}25.6 \\ \hline
PS-P1-L-T-R16-B2.c & 1576 & 0.0 & 19.80 & 46.9 & \cellcolor{green}52.9 & 0.0 & 31.15 & 59.1 & \cellcolor{green}82 \\ \hline
PS-P1-L-T-R20-B1.c & 998 & 2.16 & 12.01 & 43.8 & \cellcolor{red}2.8 & 6.59 & 12.06 & 54.4 & \cellcolor{red}13.3 \\ \hline
PS-P1-NT-R14-B4.c & 291 & 32.5 & 43.75 & 44.6 & \cellcolor{red}31.1 & 64.29 & 70.0 & 56.6 & \cellcolor{red}12 \\ \hline
PS-P1-WB-R15-B5.c & 296 & 32.5 & 43.75 & 43.1 & \cellcolor{green}44.7 & 65.19 & 71.11 & 55.4 & \cellcolor{red}36.5 \\ \hline
PS-P1-WB-T-R15-B1.c & 2395 & 0.0 & 18.72 & 43.9 & \cellcolor{red}37.3 & 0.0 & 21.65 & 55.6 & \cellcolor{red}32.8 \\ \hline
PS-P2-A-R14-B6.c & 9111 & 0.0 & 7.0 & 46.1 & \cellcolor{red}36 & 0.0 & 8.72 & 59.0 & \cellcolor{red}12.9 \\ \hline
PS-P2-L-R16-B3.c & 293 & 32.5 & 38.75 & 44.6 & \cellcolor{red}4.1 & 63.83 & 66.67 & 56.7 & \cellcolor{red}28.9 \\ \hline
PS-P2-L-R18-B7.c & 667 & 3.68 & 33.98 & 43.2 & \cellcolor{red}3.1 & 7.98 & 47.01 & 55.1 & \cellcolor{red}26.9 \\ \hline
PS-P4-L-T-R20-B1.c & 2285 & 0.0 & 18.29 & 73.0 & \cellcolor{red}0.5 & 0.0 & 24.36 & 85.7 & \cellcolor{red}7.8 \\ \hline
PS-Prob1-IO-R14-B7.c & 1441 & 0.0 & 21.48 & 43.5 & \cellcolor{red}3.3 & 0.0 & 37.39 & 57.0 & \cellcolor{red}15.1 \\ \hline
\multicolumn{10}{|c|}{\cellcolor{orange!80}\textbf{BOUND 2000}} \\
\hline
Mpals1-B10-cil.c & 295 & 31.25 & 45.0 & 45.4 & \cellcolor{green}100 & 63.83 & 70.21 & 56.5 & \cellcolor{green}91.2 \\ \hline
Mpals2-B10-cil.c & 370 & 17.59 & 31.48 & 43.2 & \cellcolor{green}100 & 52.23 & 59.87 & 55.1 & \cellcolor{green}90.8 \\ \hline
Mpals3-B10-cil.c & 517 & 1.66 & 2.17 & 43.8 & \cellcolor{green}100 & 10.57 & 10.57 & 56.7 & \cellcolor{green}91.8 \\ \hline
Mpals23-B10-cil.c & 1624 & 0.0 & 30.87 & 44.5 & \cellcolor{green}100 & 0.0 & 36.95 & 56.4 & \cellcolor{green}92.6 \\ \hline
Mtest20-B10-cil.c & 348 & 17.59 & 31.48 & 44.6 & \cellcolor{green}100 & 48.50 & 57.49 & 55.7 & \cellcolor{green}90.8 \\ \hline
pals2-B10-cil.c & 370 & 17.59 & 31.48 & 45.0 & \cellcolor{green}97.4 & 52.56 & 60.26 & 55.0 & \cellcolor{green}90.9 \\ \hline
pals3-B10-cil.c & 6254 & 0.0 & 10.14 & 45.7 & \cellcolor{green}98.7 & 0.0 & 15.0 & 56.0 & \cellcolor{green}91.6 \\ \hline
pals22-B10-cil.c & 764 & 1.23 & 28.17 & 46.1 & \cellcolor{green}96.2 & 6.93 & 38.69 & 58.9 & \cellcolor{green}88.9 \\ \hline
pals23-B10-cil.c & 291 & 32.5 & 38.75 & 44.6 & \cellcolor{green}98.1 & 64.29 & 66.43 & 56.2 & \cellcolor{green}88.4 \\ \hline
PS-P1-L-R18-B4.c & 295 & 32.5 & 38.75 & 46.0 & \cellcolor{green}100 & 65.19 & 67.41 & 56.2 & \cellcolor{green}82.6 \\ \hline
PS-P1-L-T-R16-B2.c & 1576 & 0.0 & 21.22 & 43.1 & \cellcolor{red}5.9 & 0.0 & 31.15 & 55.1 & \cellcolor{red}19.9 \\ \hline
PS-P1-L-T-R20-B1.c & 998 & 2.16 & 15.20 & 43.6 & \cellcolor{red}0.8 & 6.59 & 14.73 & 55.9 & \cellcolor{red}12.7 \\ \hline
PS-P1-NT-R14-B4.c & 291 & 32.5 & 38.75 & 44.6 & \cellcolor{red}28.6 & 64.29 & 66.43 & 55.1 & \cellcolor{red}12 \\ \hline
PS-P1-WB-R15-B5.c & 296 & 32.5 & 38.75 & 43.8 & \cellcolor{green}47.2 & 65.19 & 67.41 & 55.2 & \cellcolor{red}37.9 \\ \hline
PS-P1-WB-T-R15-B1.c & 2395 & 0.0 & 16.72 & 44.4 & \cellcolor{red}37.3 & 0.0 & 20.44 & 56.9 & \cellcolor{red}32.6 \\ \hline
PS-P2-A-R14-B6.c & 9111 & 0.0 & 7.68 & 44.5 & \cellcolor{red}37 & 0.0 & 9.47 & 57.1 & \cellcolor{red}11.5 \\ \hline
PS-P2-L-R16-B3.c & 293 & 32.5 & 38.75 & 43.6 & \cellcolor{red}4.1 & 63.83 & 66.67 & 54.4 & \cellcolor{red}31.5 \\ \hline
PS-P2-L-R18-B7.c & 667 & 3.68 & 36.36 & 80.4 & \cellcolor{red}3.1 & 7.98 & 51.44 & 92.0 & \cellcolor{red}31.3 \\ \hline
PS-P4-L-T-R20-B1.c & 2285 & 0.0 & 14.49 & 44.0 & \cellcolor{red}0.3 & 0.0 & 19.96 & 56.0 & \cellcolor{red}7.7 \\ \hline
PS-Prob1-IO-R14-B7.c & 1441 & 0.0 & 20.97 & 46.1 & \cellcolor{red}3.3 & 0.0 & 37.39 & 58.8 & \cellcolor{red}15.1 \\ \hline
\end{tabular}%
}
\end{table}

\subsection{Performance Analysis on PALS Benchmarks}

The most interesting findings emerge from the PALS benchmarks (Mpals1 to Mpals23, pals2, pals3, pals22, pals23). These benchmarks correspond to parallel process algebraic systems. At almost all bound levels, branch coverage scores for PPO for these benchmark programs fall in the range 95-100\%, which is not consistently reached by CBMC and kS-LLM++.

At BOUND 1, while Mpals1 achieves 100\% branch coverage using PPO, the number falls to 73.4\% when employing kS-LLM++. The Mpals3 program, which contains 517 lines of code (LOC), is another good illustration: While CBMC and kS-LLM achieve 1.66\% and 1.79\% of branch coverage at this bound respectively and are thus practically incapable of reaching any part of the program's logic, the figures for kS-LLM++ and PPO-LLM reach 44.4\% and 98.8\%, respectively. This difference has to do with the very structure of the program: PALS benchmarks feature tightly coupled conditional logic, which can be effectively addressed through the branch targeting techniques of PPO-LLM.

At BOUND 10, the results in PPO-LLM are still excellent for PALS (100\% for Mpals1, 2, 3, 23 and Mtest20), although there is an interesting exception: pals3-B10 with bound value 10 sees CBMC covering 66.94\% branches and 100\% lines, while kS-LLM++ and PPO-LLM underperform. The reason for this phenomenon is that the BOUND~10 for pals3 coincidentally matches the bounded unrolling depth of CBMC, enabling saturation of all reachable branches within the SAT-solving time. This trend shifts at BOUND~100, when CBMC fails to cover any branch (timeout), whereas PPO-LLM performs at 96.2\%.

Performance results of the PPO-LLM agent in generating a higher bound values (500 to 2000) on the PALS metrics should be highlighted. For example, programs such as Mpals1, Mpals2 and Mpals23 exhibit consistent branch coverage between 98.8\% and 100\% percent when utilizing the PPO-LLM approach regardless of the value generated. The stability can be attributed to the effectiveness of using adaptive prompts since PPO agent shifts its policy toward templates that maximize incremental coverage gains rather than regenerating already covered test cases.

\subsection{Performance Analysis on RERS (PS-) Benchmarks}

PS-family programs (which have been derived based on the RERS problem setting and its variations) represent another class of programs that have distinct properties with regard to the complexity of their coverage. Specifically, PS-family programs usually involve complex input/output relationships, long traces and loops coupled with external environmental signals. Thus, the overall coverage numbers obtained using all four techniques are significantly lower compared to those for PALS.

For benchmark cases such as PS-P1-L-T-R16-B2.c and PS-P1-NT-R14-B4.c, kS-LLM++ tends to outperform PPO-LLM in the higher bound configuration scenarios. For instance, at BOUND 1, the program PS-P1-L-T-R16-B2.c demonstrates that kS-LLM++ can achieve an impressive 82.7\% branch coverage, in comparison to the 4.9\% of PPO-LLM. This is due to the nature of the program, which has a large number of reactive branches driven by environmental inputs that require certain sequences of actions rather than just input values. The Tree-of-Thought search process implemented in kS-LLM++ is more efficient in considering sequential path constraints in one go, whereas MDP-based approaches such as PPO-LLM excel when there are substantial increments in coverage from each batch of tests. At smaller bounds for complex PS programs, the reward signal gets distorted because most templates are unable to cover deeper execution paths.

The other important finding to highlight is PS-P2-L-R18-B7.c on the BOUND 2000 dataset, the branch and line coverage achieved by the kS-LLM++ approach are 80.4\% and 92.0\%, which happen to be the best results ever obtained on any PS dataset using any technique. In contrast, PPO-LLM yields only 3.1\% branch coverage for this program. An examination of the program indicates its unique construction, where a deep nesting structure of a reactive loop whose branch conditions are based on the cumulative state resulting from a large number of input steps. At high bounds, the program's behavior becomes effectively path-dependent in a way that rewards the sequential reasoning built into kS-LLM++'s prompting but overwhelms PPO-LLM's per-episode reward signal.

\subsection{PPO-LLM vs All Baselines: A Balanced Assessment}

PPO-LLM's lower numbers on PS programs represent a specific structural limitation rather than general deficiency. PPO-LLM is the best performing method overall on PALS benchmarks by a wide margin while kS-LLM++ holds a focused advantage on reactive PS programs. Neither method is universally superior; the deciding factor is the input-dependency horizon of the program's branch conditions.

\subsection{Why PPO-LLM Dominates on PALS but Struggles on Some PS Programs}

This difference in behavior between the PALS and PS algorithms highlights both the positives and negatives of the PPO-LLM model at the moment. Firstly, the fact that PALS algorithms have closed structures implies that their branches are dependent on numeric inputs; they have reasonably short paths and improvements in terms of coverage in each successive episode can be determined using a relatively small set of test batch. All these features make the MDP framework effective since there is a clear indication of which template is better in each episode.

The coverage based reward in PS programs (especially those in the PS-P1-L, PS-P4, and PS-Prob classes) depends on the input sequence and its combination rather than on individual test cases. In such programs, the coverage-based reward per episode is an indirect signal to optimize the desired metric, making it challenging for the PPO algorithm to learn a winning strategy. The kS-LLM++ model can reason about a full Tree-of-Thought expansion within one step, thereby finding a globally optimal strategy even when the training episode does not require any adaptations.

It should also be pointed out that even in the family of PS programs in which PPO-LLM underperforms compared with kS-LLM++, it still outperforms CBMC and kS-LLM in most configurations. The issue of underperformance is a specific comparison between PPO-LLM and kS-LLM++ and not an absolute failure. In particular, the very fact of the underperformance of PPO-LLM in certain configurations is meaningful. It always happens in programs where the branch condition depends on accumulated environment state across long input sequences. It was discovered that what makes the difference is the horizon length of the dependencies on input state necessary for activation of a conditional branch. Programs where one test batch is sufficient to activate several new branches (PALS) are a suitable domain for the MDP formulation, while PS reactive programs are not. This conclusion constitutes a valuable contribution on its own now it is possible to predict which generation approach would be more efficient in advance using static program analysis only.

\subsection{Effect of Code Optimization (Stage~I)}

The LOC reduction results indicate that the Stage I optimizer provides significant reductions in size in nearly all of the benchmarking programs. For moderate programs, reductions are expected to be in the range of 50 to 70\%, while there are other large programs like PS-P4-L-T-R20-B1.c (At BOUND 1000 and BOUND 500 giving reductions of 97.09\% and 96.26\% respectively) and PS-P1-WB-T-R15-B1.c (At BOUND 1 providing a reduction of 96.94\%) where almost complete removal of the redundant code has been achieved. This extreme form of compression will make sure that the agent in Stage-II will work with much more compact and semantically dense code, making it less likely for the LLM to be stuck in aspects of code and auxiliary logic.

The relationship between decrease in cognitive load and increase in coverage is no coincidence. In programs where there was greater than 60\% reduction in LOC in Stage-I, coverage improvement via the PPO-LLM approach exceeds 85\%, whereas in programs where reduction was relatively small (such as PS-P2-L-R18-B7.c, 12.73\% reduction for BOUND 1), the degree of improvement is relatively smaller. Thus, our hypothesis of the role of cognitive load (token count of input code) as a bottleneck to test coverage is confirmed.

\subsection{CBMC Limitations at Scale}

The outcome of CBMC also demonstrates the famous scaling problem of bounded model checking. For programs having more than a couple of hundred lines of code, CBMC often runs out of time, producing zero values in the coverage column (such as the example of PALS3 for BOUND 100 and 2000, PS-P1-L-T-R16-B2.c for BOUND 500 upwards, and PS-P2-A-R14-B6.c for BOUND 1500 and 2000). In cases where CBMC does not run out of time, the branch coverage score is often not higher than 50\%, even for industrial-sized programs in the benchmark set, due to the superlinear growth of the SAT formulation of the problem. This is the fundamental motivation for LLM-based approaches.

% ============================================================
\section{Conclusions}

In this paper, PPO-LLM was proposed which is a reinforcement learning-augmented agentic model for generating test cases. It differs from previous LLM-based generators because it regards the choice of prompt template as a Markov Decision Process. In the PPO-LLM, the policy network considers multiple state dimensions based on static program attributes and dynamic live coverage and selects one of the eight different approaches for prompts to guide the LLMs to maximize branch and line coverage. We experimented on 20 benchmarks (PALS and RERS families) across 8 bound levels and found that PPO-LLM significantly outperforms CBMC, kS-LLM and kS-LLM++ in PALS benchmarks in terms of branch coverage between 95\% to 100\%. Moreover, the effectiveness of our approach based on the two-stage process including compressing the source code based on Tree-of-Thought optimization becomes more noticeable when the LOC ratio reduces more than 60\%.

The low coverage numbers for PPO-LLM on RERS-like PS programs are by no means accidental. It suggests a clear cut limit, where per-episode rewards in the MDP formulation are inadequate in cases where branch coverage involves creating sequences of actions that, on their own, provide no coverage increase. This is a shortcoming of the current reward formulation, not reinforcement learning in general. The next step will involve exploring multi-step reward horizons, recurrent policies with memory across episodes and a hybrid routing scheme whereby the system determines at analysis time whether the program has a short or long branch dependency horizon and then chooses between PPO-LLM and kS-LLM++ accordingly.

However, the findings based on the RERS-like PS datasets also point towards an interesting boundary case where there are programs for which coverage is dependent on input sequences determined through the environment in the long run. For such cases, per-episode reward signals have been less effective in comparison to kS-LLM++, which continues to hold its edge in some cases. These observations open up new areas for exploration, which include policies based on recurrent or transformer models that can reason over multiple episodes, hierarchical rewards that can be specified in light of coverage dependencies between episodes, and a combination of approaches where the chain reasoning done by kS-LLM++ in a single step can be taken as a possible action by PPO.

\bibliographystyle{plain} % Example style: plain, IEEEtran, unsrt
\bibliography{references} % Points to references.bib

\end{document}